\documentclass[prl,twocolumn,preprintnumbers,amsmath,amssymb]{revtex4}
\usepackage{epsfig,psfrag}

\newcommand{\insertfig}[2]{\mbox{\epsfxsize=#1cm \epsfbox{#2.eps}}}

\begin{document}

\title{Pomeron dominance in deeply virtual Compton scattering \\
 and the femto holographic image of the proton }
\author{D. M\"uller}

\affiliation{Department of Physics and Astronomy, Arizona State
University, Tempe, AZ 85287-1504, USA}

\begin{abstract}
\noindent
The dominance of the soft pomeron in soft high energy
scattering and the  evolution to the deeply virtual
regime, predicted by perturbation theory, allow us to
reveal generalized parton distributions from H1 and ZEUS
measurements of deeply virtual Compton scattering. These
distributions encode a holographic image of the proton,
which will be presented.
\end{abstract}

\maketitle

We study deeply virtual Compton scattering (DVCS) off the proton
at high energies and aim to reveal generalized parton
distributions (GPDs) \cite{MueRobGeyDitHor94Ji96Rad96} from HERA
measurements \cite{Adletal01Chekanov:2003yaAktas:2005ty}. The GPDs
then yield a 3D image of quark and gluon distributions at small
longitudinal momentum fraction $x$. The analysis substantially
relies on the soft pomeron dominance in the non-perturbative
sector. Various points of view on the pomeron are recalled before
the DVCS predictions are derived in a simple, however,
approximated analytic form and the analysis is performed.

\begin{figure*}[th]
\begin{center}
\mbox{
\begin{picture}(600,70)
\put(-2,-2){\insertfig{4}{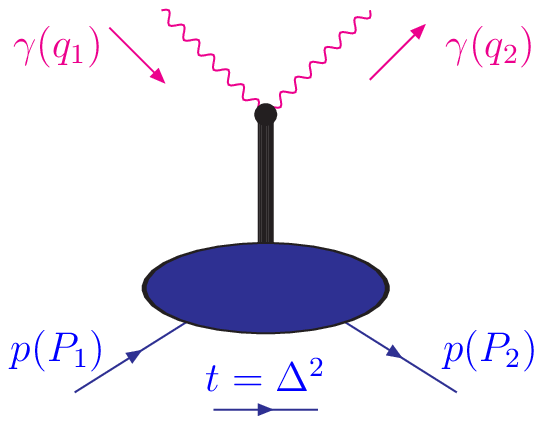}}
\put(50,-12){(a)}
\put(115,0){\insertfig{5.1}{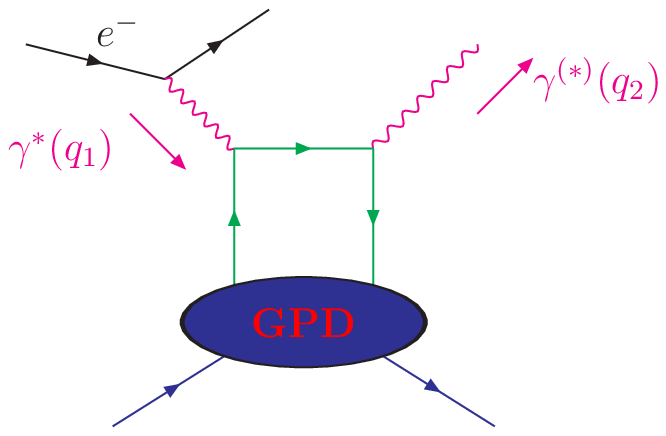}}
\put(177,-12){(b)}
\put(270,0){\insertfig{3}{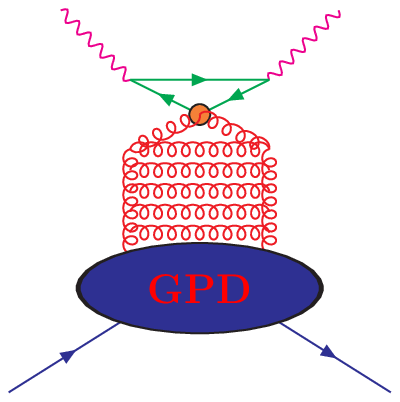}}
\put(310,-12){(c)}
\put(375,2){\insertfig{4.5}{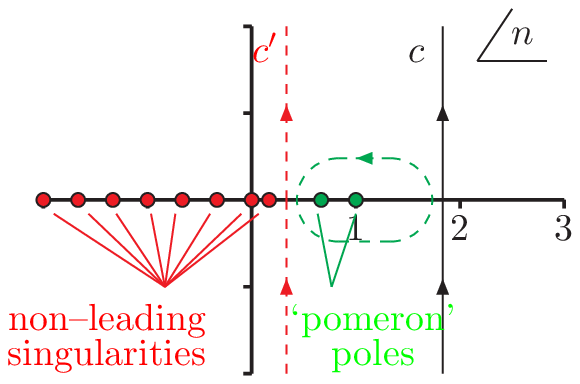}}
\put(432,-12){(d)}
\end{picture}
}
\end{center}
\caption{\label{Fig-CS} The Compton scattering amplitude (a) at
high energy and small $t$ is dominated by the  pomeron exchange in
the $t$--channel. The lowest order perturbative contribution to
DVCS at the scale ${\cal Q}^2=-q_1^2$ and at a very low input
scale is shown in (b) and (c), respectively. The integration path
used in Eq.\ (\ref{Def-CFF}) and its deformation in the complex
$n$--plane are shown in (d). }
\end{figure*}

In soft high energy scattering experiments, i.e., no further large
scale is present, the energy dependence of scattering  amplitudes
is described within Regge phenomenology as an exchange of
particles in the $t$ (momentum transfer squared) channel, see
Fig.\ \ref{Fig-CS}(a). Theoretical considerations, based on
partial wave decomposition, analyticity, and   crossing, predict
that scattering amplitudes behave at large center--of--mass energy
squared $s$ as $s^{\alpha(t)}$, where $\alpha(t)=\alpha(0) +
\alpha^\prime t$ is the leading Regge trajectory in terms of the
intercept $\alpha(0)$ and slope $\alpha^\prime$. Such a trajectory
relates the particle spin $J=\alpha(m^2)$ with its mass. To
explain experimental measurements, the trajectory
$\alpha_{\mathbb{P}}(t)=1.08 + 0.25 t$ has been introduced
\cite{Donnachie:1984xqDonnachie:1985izDonnachie:1987pu}. It is
associated with a hypothetical spin--one quasi particle, the
so--called soft pomeron that carries the vacuum quantum numbers.
No particles belonging to this pomeron trajectory could be
established and so it remains exceptional.

The notion pomeron is also applied  in the phenomenology of (semi)
hard reactions. The HERA measurements, where a 27 GeV
electron/positron beam collides with a 820 GeV proton one, lead to
the discovery that the energy dependence of cross sections get
steeper with growing photon virtuality $Q^2=-q_1^2$. Let us
consider the proton structure function $F_2(x_{\rm Bj},Q^2)$,
measured in deep inelastic scattering (DIS), at small $x_{\rm
Bj}$. The reciprocal Bjorken scaling variable $1/x_{\rm
Bj}=2P_1\cdot q_1/Q^2$, where $P_1$ is the proton momentum, is
nearly proportional to the photon--proton center--of--mass energy
squared $W^2=(q_1+P_1)^2$. $F_2$ possesses an almost flat
$1/x_{\rm Bj}$ dependence at low $Q^2 <  Q_0^2 \sim 1 {\rm GeV}^2$
and gets steeper with increasing  $Q^2$:
\begin{equation}
\label{F2-beh}
F_2(x_{\rm Bj},Q^2) \sim (1/x_{\rm Bj})^{\lambda(Q^2)}\,,
\end{equation}
where $\lambda(Q^2) \gtrsim 0$ rises with growing $Q^2$.

In Regge theory $\lambda = \alpha_{\mathbb{P}}(0)-1$ is considered
as $Q^2$ independent, which is incongruous with experiment. To
cure this issue, a double pomeron exchange within a soft and
hard pomeron, e.g., $\alpha_{\mathbb{P}}^{\rm hard}(0)\approx
1.4$, was suggested \cite{Donnachie:1998gmDonnachie:2001xx}. The
(energy ordered) resummation of the gluon ladder in the BFKL
approach results the perturbative pomeron, which indeed possesses
such a large intercept \cite{BFKL}. However, the theoretical
understanding remains poor.

%%%%%%%%%%%%%%%%%%%%%%%%%%%%%%%%%%%%%%%%%%%%%%%%%%%%%%%%%%%%%%%%%%%%%
%            Figure
%%%%%%%%%%%%%%%%%%%%%%%%%%%%%%%%%%%%%%%%%%%%%%%%%%%%%%%%%%%%%%%%%%%%%
%%%%%%%%%%%%%%%%%%%%%%%%%%%%%%%%%%%%%%%%%%%%%%%%%%%%%%%%%%%%%%%%%%%%%

Fortunately, perturbative QCD predicts the variation of
observables with respect to a hard scale. $F_2$ is the absorptive
part of the forward Compton amplitude. Analogous to Fig.\
\ref{Fig-CS}(b), it factorizes in terms of quark $q(x,Q^2)$ and
gluon $G(x,Q^2)$ densities. Their scale changes are governed by
evolution equations, written for the Mellin--moments, e.g.,
$G_n(Q^2)=\int_0^1 dx x^{n-1} G(x,Q^2)$, as
\begin{eqnarray}
\label{Def-EvoEqu}
Q^2\frac{d}{dQ^2} \left(\Sigma_n \atop G_n\right) = \left({
{^{\Sigma\Sigma}\!\gamma_n} ^{\Sigma G}\!\gamma_n} \atop
{^{G\Sigma}\!\gamma_n} {^{G G}\! \gamma_n}  \right) \left(\Sigma_n
\atop G_n\right)\,.
\end{eqnarray}
Here $\Sigma_n=\sum_q q_n$ is the  flavor singlet quark
density and ${^{AB}\!\gamma_n}$ are the anomalous dimensions,
depending on spin $n$. The gluon--gluon and gluon--quark
anomalous dimensions possess a `pomeron' pole at spin $n=1$.
Solving the evolution equation yield an exponentiation of this
pole.  In  $x$--space  this converts  to the so--called double log
approximation, see  below. This resummation of large
contributions might be effectively  parameterized as  in
(\ref{F2-beh}).

The DIS measurements are well described by perturbative QCD, where
the input parton densities  are extracted from global fits. It has
been suggested to generate dynamically  sea quarks and gluon
densities by evolution \cite{GluReyVog90}. Starting from a very
low input scale, fine tuning is required to fit experimental data.
In the small $x$ region the evolution is mainly driven by the
`pomeron' pole of the anomalous dimensions and data are indeed
consistent with the double log approximation \cite{Ball:1994kc}.
Both approaches were combined  within a soft pomeron like input
\cite{Kotikov:1998qt}, where only the initial scale, two
normalization factors,
\begin{eqnarray}
\label{Ans-SofPom}
\Sigma_n(Q_0) = \frac{N_\Sigma}{n-\alpha_{\mathbb{P}}(0)}\,,\qquad
G_n(Q_0) = \frac{N_G}{n-\alpha_{\mathbb{P}}(0)}\,,
\end{eqnarray}
and perhaps the running coupling have to be adjusted.

The dynamical generation of parton densities for small $x_{\rm
Bj}$ is perhaps the most predictive approach to DIS. Although the
`pomeron' pole in the evolution operator is essential, the
resulting parton densities depend on the non--perturbative input,
too. Since real Compton scattering can be described at high energy
within  soft pomeron exchange \cite{Aid:1995bzChekanov:2001gw}, it
is apparent that the non-perturbative input is dominated at a low
scale  by the soft pomeron, too. It arises in the expectation
value of operators, sandwiched between proton states, as pole
contribution (\ref{Ans-SofPom}).

Let us apply this conjecture to DVCS. At leading order (LO) the
factorization of the amplitude is given by the hand--bag diagram,
depicted in Fig.\ \ref{Fig-CS}(b), in terms of Compton form
factors (CFFs) \cite{BelMueKir01}. At small $\xi\cong x_{\rm
Bj}/2$ only the flavor singlet part of the helicity conserved and
charge even  CFF ${\cal H}(\xi,t,{\cal Q}^2=-q_1^2)$ is relevant
\cite{footnote}. To diagonalize the evolution operator, the
complex conformal partial wave decomposition will be employed
\cite{MueSch05,ManKirSch05}
\begin{eqnarray}
\label{Def-CFF}
{\cal H}^{\rm S}%(\xi,t,{\cal Q}^2)
&\!=\!& \frac{1}{2 i}
\int_{c-i\infty}^{c+i\infty} dn
\frac{\Gamma(n+3/2)}{\Gamma(3/2)\Gamma(2+n)}
 \left[i - \cot\left(\frac{\pi n}{2}\right) \right]
%\frac{2 i}{1-e^{i n \pi }}
\\
&&\times\left(\frac{2}{\xi}\right)^{n} \left(Q_S^2,0\right) \mathbb{E}_n({\cal Q}^2,{\cal
Q}^2_0) \left({\Sigma_n \atop G_n}\right)(\xi,t,{\cal Q}_0^2)\,.
\nonumber
\end{eqnarray}
Here the singularities in the complex $n$-plane are on the l.h.s.\
of the integration path, cf.\ Fig.\ \ref{Fig-CS}(d). $Q_S^2=
\sum_q Q_q^2/n_f$ is the averaged  charge square fraction for
$n_f$ active quarks. $\Sigma_n$ and $G_n$ are conformal GPD
moments, which reduce in the forward case to the parton density
ones. The evolution operator $\mathbb{E}_n$, a $2\times2$ matrix,
is the same as in DIS, see, e.g., Ref.\ \cite{Kotikov:1998qt}. By
means of $\mathbb{P}^\pm_n $ projectors it might be  decomposed in
`$+$' and `$-$'  eigenmodes:
\begin{eqnarray}
\label{Def-EvoOpe}
\mathbb{E}_n({\cal Q}^2,{\cal Q}^2_0) =\mathbb{P}^+_n
\left(\frac{\alpha_s({\cal Q})}{\alpha_s({\cal
Q}_0)}\right)^{\lambda^+_n} + \mathbb{P}^-_n
\left(\frac{\alpha_s({\cal Q})}{\alpha_s({\cal
Q}_0)}\right)^{\lambda^-_n}\,,
\end{eqnarray}
where $\lambda^\pm_n $ are eigenvalues of the anomalous dimension
matrix, appearing in Eq.\ (\ref{Def-EvoEqu}). The running coupling
\begin{eqnarray}
\alpha_s({\cal Q}) = \frac{4\pi}{(11 - 2n_f/3 )
\ln({\cal Q}^2/\Lambda_{\rm QCD}^2)}
\end{eqnarray}
is parameterized by $\Lambda_{\rm QCD}$.

%%%%%%%%%%%%%%%%%%%%%%%%%%%%%%%%%%%%%%%%%%%%%%%%%%%%%%%%%%%%%%%%%%%%%
%            Figure
%%%%%%%%%%%%%%%%%%%%%%%%%%%%%%%%%%%%%%%%%%%%%%%%%%%%%%%%%%%%%%%%%%%%%
\begin{figure*}[t]
\begin{center}
\mbox{
\begin{picture}(800,90)
\psfrag{ds}[cc][cc]{$\frac{d\sigma}{dt}(W,t,{\cal Q}^2)$}
\psfrag{t}{\hspace{21mm}${ \atop -t}$}
%\psfrag{a}{\hspace{2cm}(a)}
\put(0,0){\insertfig{5.8}{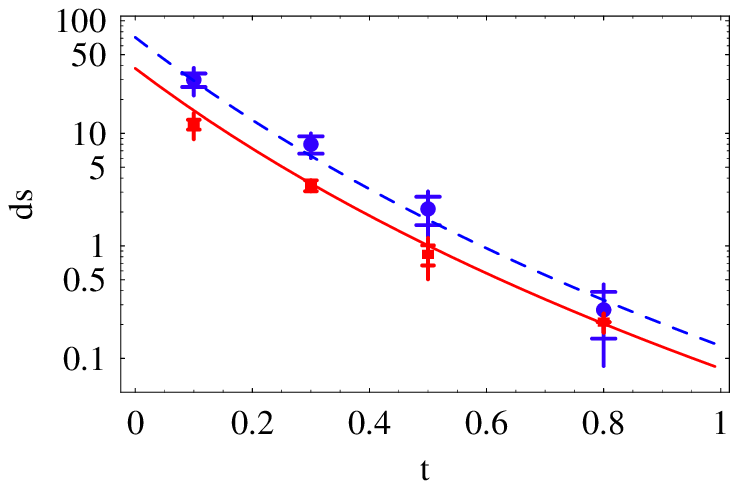}}
\psfrag{cQ2}{\hspace{23mm}${ \atop {\cal Q}^2}$}
%\psfrag{b}{\hspace{2cm}(b)}
\psfrag{si}[cc][cc]{$\int_0^{|t_{\rm cut}|} \!d|t|\,\frac{d\sigma}{dt}(W,-|t|,{\cal Q}^2)$}
\put(180,-2){\insertfig{5.85}{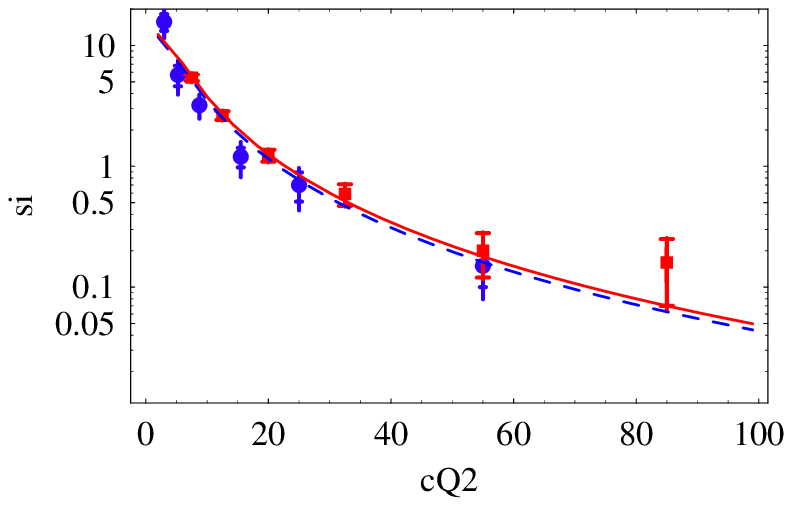}}
\psfrag{W}{\hspace{22mm}${ \atop W}$}
%\psfrag{c}{\hspace{2cm}(c)}
\put(350,0){\insertfig{5.8}{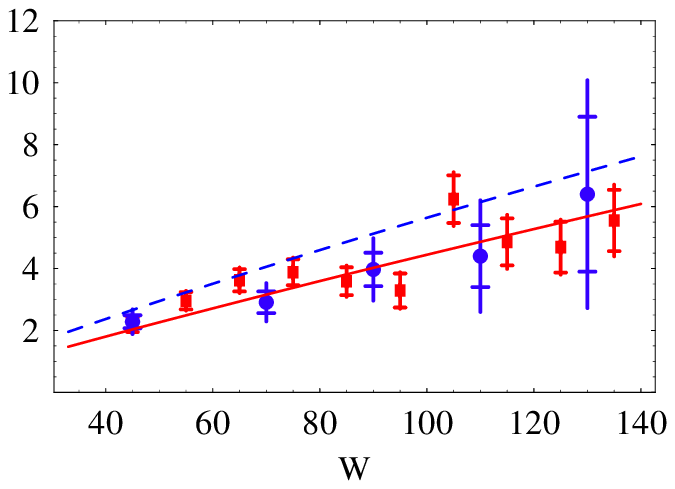}}
\end{picture}
}
\end{center}
\caption{In the left panel the $t$--dependence of the
differential DVCS cross section (\ref{Def-CroSec}) is fitted to
H1 data  for $W=71\,{\rm GeV},\; {\cal Q}^2=4\,{\rm GeV}^2$
(circles, dashed) and $W=82\,{\rm GeV},\; {\cal Q}^2=8\,{\rm
GeV}^2$ (squares, solid). The integrated cross section  with
$t_{\rm cut}=-1 {\rm GeV}^2$ versus ${\cal Q}^2$ for $W=82 [89]\,
{\rm GeV}$ (middle) and  versus $W$ for ${\cal Q}^2=8 [9.6]{\rm GeV}^2$
(right) are compared with H1 (circles, dashed) [ZEUS (squares,
solid)] data \cite{Adletal01Chekanov:2003yaAktas:2005ty}. The
statistical and systematic errors are added in quadrature.
\label{Fig-Data} }
\end{figure*}
%%%%%%%%%%%%%%%%%%%%%%%%%%%%%%%%%%%%%%%%%%%%%%%%%%%%%%%%%%%%%%%%%%%%%

To illuminate the physics at small $x_{\rm Bj}$, let us derive the
double log approximation for ${\cal H}^{\rm S}$ (\ref{Def-CFF}).
The dominant contribution comes from the most right--lying
singularity, see Fig.\ \ref{Fig-CS}(d). If $\alpha_{\mathbb{P}}(t)
<1$, this is an essential singularity appearing in the `$+$'--mode
of the evolution operator (\ref{Def-EvoOpe}),
\begin{eqnarray}
{\lambda_n^+} = -\frac{\lambda}{n-1} + \bar{\lambda}^+ +O(n-1)\,,
\quad \lambda=\frac{36}{33 - 2 n_f}\,.
\end{eqnarray}
For $0<\alpha_{\mathbb{P}}(t)$ the next--leading singularity  is
expected to be the soft pomeron pole. To simplify the analysis, we set
$\alpha_{\mathbb{P}}(0)=1$. Hence, for $t=0$ this singularity
joins the leading one and both of them should be considered in
Eq.\ (\ref{Def-CFF}). They can be encircled   by a deformation of
the integration contour,  as shown in Fig.\ \ref{Fig-CS}(d). The
remaining background integral (dashed integration path $c^\prime$)
can be safely neglected. Next the variable transformation $n=1+
\rho j$ with $\rho= \left[\lambda\ln(\alpha_s({\cal
Q}_0)/\alpha_s({\cal Q}))/\ln(2/\xi)\right]^{1/2}$ and ${\cal Q} >
{\cal Q}_0$ is performed, where $\rho$ is considered as small.
Taylor expansion at $\rho=0$ yield integrals that reduce for $t=0$
to the modified Bessel function $I_a(\sigma)$ with $a=0,1,\cdots$:
\begin{equation}
\label{Def-Int-I}
I_a(\sigma, \bar{\alpha} t) = \frac{1}{2 i \pi}
\oint_{(0,\bar{\alpha} t)}\!dj\, \frac{j^a}{j-\bar{\alpha} t}
e^{\sigma (j + 1/j)/2}\,,
\end{equation}
where $\sigma^2= 4\lambda \ln[\alpha_s({\cal Q}_0)/\alpha_s({\cal
Q})] \ln[2/\xi]$. The effective slope
$\bar{\alpha}=\alpha^\prime_{\mathbb{P}}/\rho$ depends on $\xi$
and $\cal Q$. Note that the asymptotic behavior of the integral
(\ref{Def-Int-I}) is $a$ independent:
\begin{equation}
\label{Def-Int-I-asy}
I_a(\sigma, \bar{\alpha} t) =
\frac{1}{1- \bar{\alpha} t}
 \frac{e^\sigma}{\sqrt{2 \pi \sigma}}\quad \mbox{for}\;\sigma\to\infty\,.
\nonumber
\end{equation}

Now the  GPD moments have to be specified. In the presence of a
hard scale the partonic content of the pomeron can be resolved
{\cite{IngSch85}}. It is expected that  the pomeron is mainly made
of glue, see e.g., Ref.\ \cite{Derrick:1995tw}.  For simplicity
the  quark distribution, including the valence quarks, will be
neglected in the small $x$ region. Finally, the CFF
(\ref{Def-CFF}) reads in the double log approximation as
\begin{eqnarray}
\label{Res-CFF-DLA}
{\cal H}^{\rm DL}%(\xi,t,{\cal Q}^2)
&\!=\!& \frac{i \pi n_f Q_S^2}{6 \xi}  \Bigg[\rho\,
I_{1}(\sigma, \bar{\alpha} t) \left(\frac{\alpha_s({\cal Q})}{\alpha_s({\cal
Q}_0)}\right)^{\bar{\lambda}^+} G_1(\xi,t,{\cal Q}_0^2) \nonumber\\
&&\qquad +\, \mbox{pole term}\Bigg]  + O(\rho^2) \,,
\end{eqnarray}
where $\bar{\lambda}^+ =1 +20 n_f/9 (33-2 n_f)$.  Here the pole
term comes from the soft pomeron  in the `$-$' mode. It is an
unsubstantial contribution that vanishes for $t=0$. The real part
of ${\cal H}^{\rm S}$ is induced by subleading terms in  $\rho$
and will not be considered here. We remark that the optical
theorem $F_1 =   {\Im}{\rm m} {\cal H}/(2\pi)\big|_{q_2\to q_1}$
and the Callan--Gross relation $F_2 = 2x_{\rm Bj} F_1$, valid to
LO accuracy, leads to the known double log approximation in DIS:
\begin{eqnarray}
\label{F2-pome-ansatz}
F_2(x_{\rm Bj},Q^2) \approx \frac{4x_{\rm Bj} }{3\pi}
{\Im}{\rm m} {\cal H}^{\rm DL}(2x_{\rm Bj},t=0,2Q^2 )\,
\end{eqnarray}
(the factor two in scale dependence is explained by \cite{footnote}).

The interpretation of Eq.\ (\ref{Res-CFF-DLA}) is depicted in
Fig.\ \ref{Fig-CS}(c), where the evolution operator, i.e., the
($k_\perp$ ordered) resummation of the gluonic ladder, is
inserted in the handbag diagram. Only the  moment with
$n=1$ is needed as non-perturbative input. It might be interpreted
as  a non-local gluon operator with dimension three and spin one
that is sandwiched between the proton states. Since in the
light--cone gauge this operator formally reduces to a local
operator, we might argue that its matrix element is
$\xi$ independent. In accordance with quark counting rules,
predicting the power fall at large $|t|$, a simple multipol ansatz
with unknown slope $B_G$ is used:
\begin{eqnarray}
\label{Ans-ConMomG}
G_1(\eta,t,{\cal Q}_0^2) \equiv G_1(t)  = \frac{N_G}{(1-B_G t/3)^3}\,.
\end{eqnarray}

The DVCS process is accessible via the photon leptoproduction and
interferes with the Bethe--Heitler Brems\-strahlungs one. For
small $x_{\rm Bj}$ the DVCS cross section dominates,  the
interference is negligible (after integration over the azimuthal
angle), and the Bethe--Heitler cross section, known in terms of
the elastic proton form factors, can be subtracted
\cite{Adletal01Chekanov:2003yaAktas:2005ty}. Theoretically, the
DVCS cross section for small $x_{\rm Bj}$ is safely approximated
by \cite{BelMueKir01}
\begin{eqnarray}
\label{Def-CroSec}
\frac{d\sigma_T}{dt}(W,t,{\cal Q}^2) \approx
\frac{4   \pi \alpha^2 }{\text{$\mathcal{Q}$}^4} |\xi\, {\cal H}^S(\xi,t,{\cal Q}^2)|^2
\big|_{\xi= \frac{{\cal Q}^2}{2 W^2+{\cal Q}^2}}\,.
\end{eqnarray}
%%%%%%%%%%%%%%%%%%%%%%%%%%%%%%%%%%%%%%%%%%%%%%%%%%%%%%%%%%%%%%%%%%%%%
%            Figure
%%%%%%%%%%%%%%%%%%%%%%%%%%%%%%%%%%%%%%%%%%%%%%%%%%%%%%%%%%%%%%%%%%%%%
\begin{figure*}[t]
\begin{center}
\mbox{
\begin{picture}(800,100)
\psfrag{0.875}{}
\psfrag{0.825}{}
\psfrag{0.775}{}
\psfrag{0.00005}{}
\psfrag{0.00002}[cc][cc]{$\;\;\;\;\;2\cdot 10^{-5}$}
\psfrag{0.00001}[cc][cc]{$10^{-5}$}
\psfrag{0.0005}{}
\psfrag{0.0002}[cc][cc]{$\;\;\;\;\;2\cdot 10^{-4}$}
\psfrag{0.0001}[cc][cc]{$10^{-4}$}
\psfrag{0.001}[cc][cc]{$10^{-3}$}
\psfrag{sqb2}[cc]{$\sqrt{\langle \mbox{\boldmath $b$}^2 \rangle}\;  [{\rm fm}]$}
\psfrag{x}{\hspace{24mm} $x$}
\put(-2,-20){\insertfig{6.9}{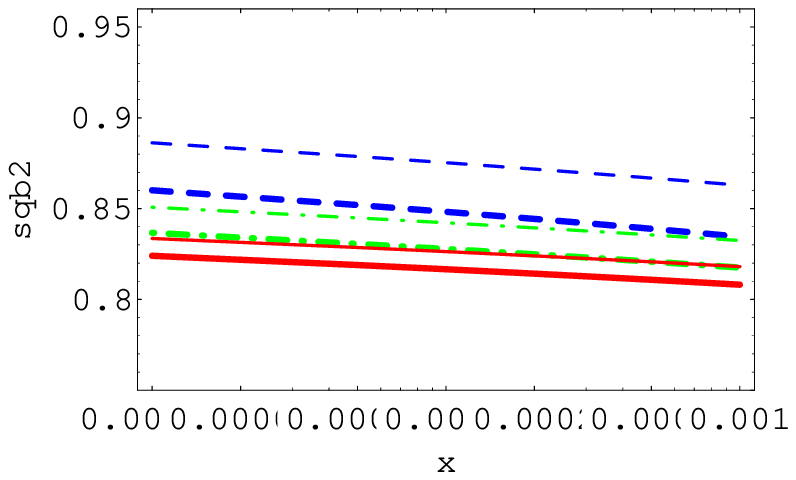}}
\psfrag{rho}[cc]{$\rho(x=10^{-3},\mbox{\boldmath $b$},{\cal Q}^2)$}
\psfrag{b}{\hspace{20mm} $|\mbox{\boldmath $b$}|\;  [{\rm fm}]$\hspace{32mm} $|\mbox{\boldmath $b$}|\;  [{\rm fm}]$}
\put(195,-14){\insertfig{6.5}{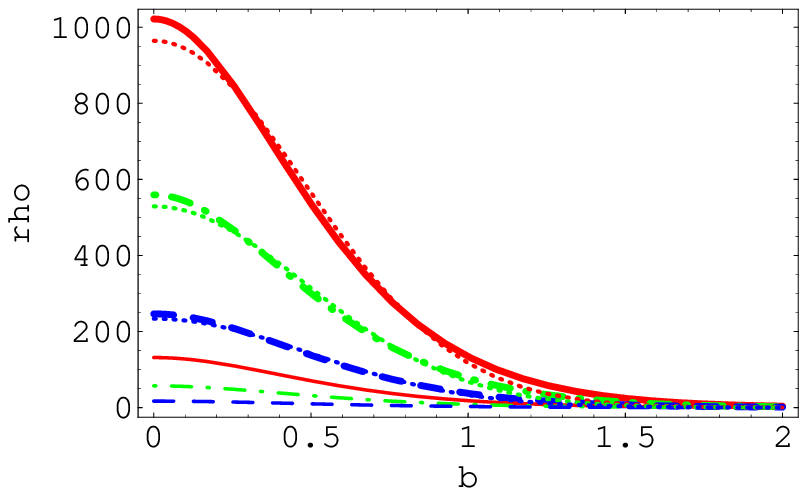}}
\put(405,0){\insertfig{3.5}{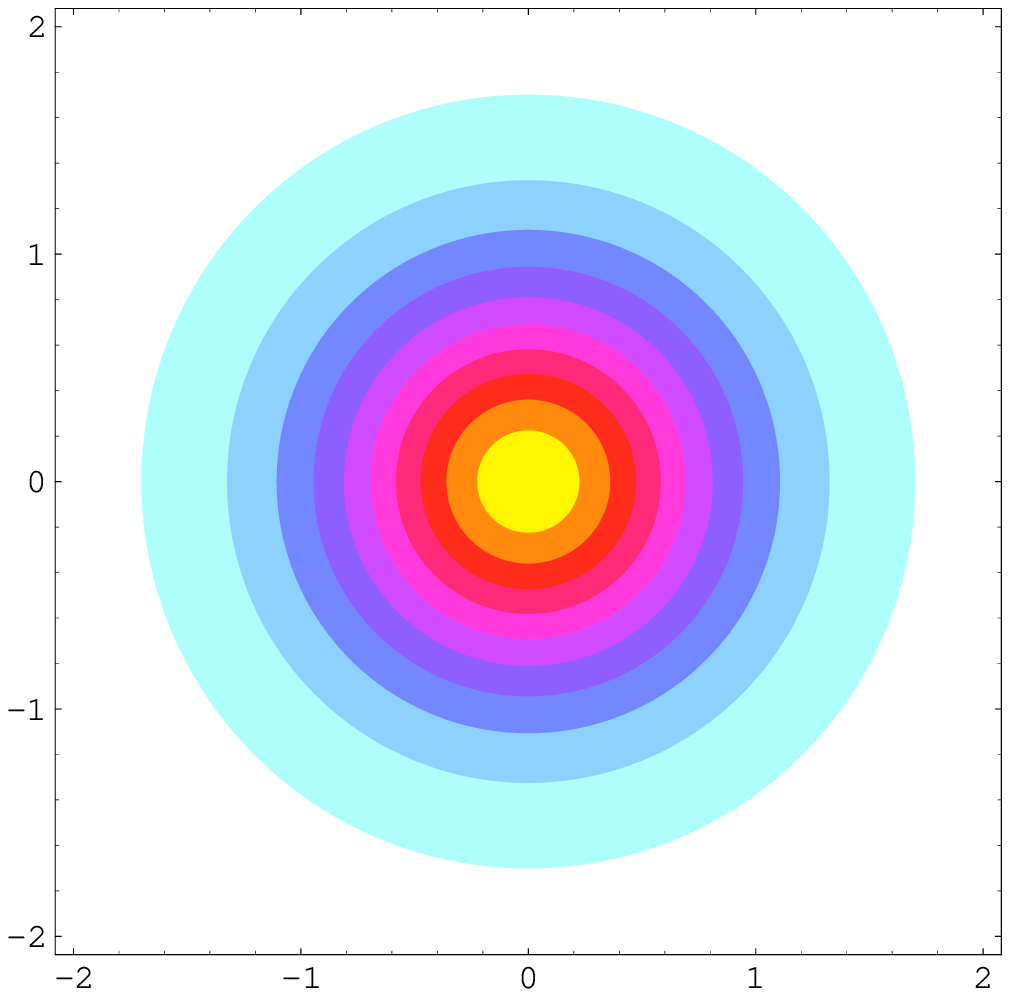}}
\end{picture}
}
\end{center}
\caption{\label{Fig-HolPic} The root of the mean square distance
(\ref{Def-b2}) for gluons (thick) and singlet quarks (thin) and
the corresponding parton densities (\ref{Def-rho})  versus
$|\mbox{\boldmath $b$}|$ are shown in the left and middle panel,
respectively,  for $x=10^{-3}$.  ${\cal Q}^2$ is set to $2\;
\mbox{GeV}^2$ (dashed), $10\; \mbox{GeV}^2$ (dash--dotted), and
$100\; \mbox{GeV}^2$ (solid). The dotted lines in the middle panel
display the result for an exponential $t$--dependence.  In the
right panel the transverse distribution of gluons is visualized,
where $x=10^{-3}$ and ${\cal Q}^2=10\; \mbox{GeV}^2$.  }
\end{figure*}
%%%%%%%%%%%%%%%%%%%%%%%%%%%%%%%%%%%%%%%%%%%%%%%%%%%%%%%%%%%%%%%%%%%%%

A fit of  the cross section, using Eqs.\ (\ref{Def-Int-I})--(\ref{Def-CroSec}),
to H1 and ZEUS data \cite{Adletal01Chekanov:2003yaAktas:2005ty} provide the
parameter set
\begin{eqnarray}
\label{ParSet}
N_G= 1.97\,,\;\;
B_G=3.68\, {\rm GeV}^{-2}\,, \;\; {\cal Q}^2_0 = 0.51\; {\rm GeV}^2\,,
\end{eqnarray}
where $n_f=3$ and $\Lambda_{\rm QCD}=150$ MeV were fixed. Note
that the  $t$-slope is rather robust under a change of
$\Lambda_{\rm QCD}$, however, it is sensitive to the form factor
shape. For an exponential ansatz it decreases to $B_G^{\rm exp}=2.58\, {\rm GeV}^{-2}$.

Let us confront these findings with DIS data in the kinematic
region that is considered here. Within the parameters
(\ref{ParSet}) the double log approximation (\ref{F2-pome-ansatz})
reasonably describes the shape of data. However, the normalization
can fail of up to 30\%. Radiative QCD corrections are important
\cite{BelMueNieSch99} and it is was stated that the normalization
discrepancy in DVCS is resolved at next--to--leading order
\cite{FreMcDStr02}. A closer look to this problem, beyond
next--to--leading order, will be presented somewhere else
\cite{Mue05aKumMueKumPasSch06}.

The dominance of the pomeron allows us now to reveal the singlet quark and
gluonic GPDs, which are represented by a  Mellin--Barnes integral
\cite{MueSch05,ManKirSch05}. In the double log approximation they read,
e.g., for vanishing skewness,  as
\begin{eqnarray}
\label{Res-GPDs-DLA}
x H^{G}(x,\eta\!=\!0)
&\!\! \simeq\!\! &  I_{0}(\sigma, \bar{\alpha} t) \left[\!\frac{\alpha_s({\cal Q})}{\alpha_s({\cal
Q}_0)}\!\right]^{\bar{\lambda}^+} \! G_1(t)
 +\cdots ,
 \\
x H^{\Sigma}(x,\eta\!=\!0)
&\!\!\simeq\!\! &  \frac{n_f}{9}\rho\, I_{1}(\sigma, \bar{\alpha} t)
\left[\!\frac{\alpha_s({\cal Q})}{\alpha_s({\cal
Q}_0)}\!\right]^{\bar{\lambda}^+}\! G_1(t)
 +\cdots\,,
 \nonumber
\end{eqnarray}
where   $\sigma(\xi=2x)$  and $\bar{\alpha}(\xi=2x)$ is set, see
below Eq.\ (\ref{Def-Int-I}). They reduce for $t=0$ to the common
parton densities.

For $\eta=0$ the Fourier transform of $H$ with respect
to the transverse momentum transfer $\mbox{\boldmath
$\Delta$}_\perp = (\mbox{\boldmath
$P$}_2-\mbox{\boldmath $P$}_1)_\perp $,
\begin{eqnarray}
\label{Def-rho}
\rho =  \int\!\!\!\!\int\!
\frac{d^2\mbox{\boldmath $\Delta$}_\perp}{(2\pi)^2}\;
e^{i\mbox{\boldmath $b$}\cdot\mbox{\boldmath $\Delta$}_\perp }
H(x,\eta=0,t=-\mbox{\boldmath $\Delta$}_\perp^2,{\cal Q}^2)\,,
\end{eqnarray}
has  a probabilistic interpretation \cite{Bur0002}. In the
infinite momentum frame the proton is viewed as a disc with a mean
square charge radius $r^2  =4 \frac{\partial}{\partial t}
F_1(t)\big|_{t=0} \approx (0.6\, {\rm fm})^2$. The density of
partons is given by $\rho(x,\mbox{\boldmath $b$},{\cal Q}^2)$,
depending on momentum fraction $x$, transverse distance
$\mbox{\boldmath $b$}$ from the proton center, and resolution
$1/{\cal Q}$. The mean square distance for a parton species is
proportional to the $t$-slope
\begin{equation}
\label{Def-b2}
\langle \mbox{\boldmath $b$}^2 \rangle =4 B\,,
\quad B= \frac{\partial}{\partial t}
\ln H(x,\eta=0,t,{\cal Q}^2)\Big|_{t=0}\,.
\end{equation}

In Fig.\ \ref{Fig-HolPic}(a) the averaged distance $\langle
\mbox{\boldmath $b$}^2 \rangle^{1/2}$ is displayed for both gluons
and sea quarks. Compared to the charge radius $r\simeq0.6$ fm,
this distance is of about 40--50\% larger. Within an exponential
form factor ansatz this inflation would shrink to 20--30\%. The
weakly grow with decreasing $x$ and fall with increasing ${\cal
Q}$ arise from the intertwining of soft--pomeron input and
evolution. The double log asymptotics, i.e., (\ref{Def-Int-I}),
(\ref{Res-GPDs-DLA}), and (\ref{Def-b2}), shed light on this:
\begin{eqnarray}
B(W,{\cal Q}^2) \simeq B_G+ \frac{\alpha_{\mathbb P}^\prime}{\sqrt{\lambda}}
\frac{\ln^{\frac{1}{2}}\left(4 W^2/{\cal Q}^2\right)}{
\ln^{\frac{1}{2}}\left(\alpha_s({\cal Q}_0)/\alpha_s({\cal Q})\right)}\,.
\end{eqnarray}
Such a behavior was also experimentally observed in hard vector
meson electroproduction, dominated by the gluon exchange.
Moreover, the measured $t$-slope of the cross section is
consistent with our findings for DVCS \cite{Adletal99}. In Fig.\
\ref{Fig-HolPic}(b) it can be realized that the sea quark density
is a  bit broader and as known from forward parton densities much
smaller than the gluon one. The gluon density within the
exponential form factor ansatz is displayed by dotted lines. For
$|\mbox{\boldmath $b$}| \lesssim 1.5$ fm the density only slightly
differs from that of the multipol ansatz. A further characteristic
is that the densities (start to) vanish for $|\mbox{\boldmath
$b$}|\gtrsim 1.5$ fm, see panel (c). However, the multipol ansatz
induces a long tail that results into an enlargement, compared to
the exponential one, of the averaged distance.

In this letter we argued  that the soft pomeron trajectory
essentially determines the gluonic GPD at a small resolution scale
and that perturbative  evolution can be used to arrive at higher
scales, where factorization holds true. This interplay of
non-perturbative and perturbative QCD leads to a realistic
behavior of the DVCS amplitude. To illuminate this, the double log
approximation has been derived and used to LO. It was checked that
a numerical treatment, with changed normalization and scale
parameter,  essentially  leads to the same results. The fit to
experimental data provides the non--perturbative parameter and
reveals so the 3D gluon and sea quark distributions inside the
proton. The theoretical framework can be refined and confronting
it with precise DVCS data would strongly constrain the
non--perturbative ansatz.

%For discussions I am indebted to A.~Belitsky.
This project has been supported by  the U.S. National Science
Foundation under grant no. PHY--0456520.

%%%%%%  Bibliography %%%%%%%%%%%%%%%%%%%%%%%%%%%%%%%%%%%%%%%%%%%%

%\bibliography{../../../../texinput/referenc,../../../../texinput/veroefli,GPDsmax}

\begin{thebibliography}{10}

\bibitem{MueRobGeyDitHor94Ji96Rad96}
D.~M{\"u}ller, D.~Robaschik, B.~Geyer, F.-M. Dittes and J.~Ho\v{r}ej{\v s}i,
\newblock Fortschr. Phys. {\bf 42}, 101 (1994);
%%CITATION = HEP-PH 9812448;%%
X.~Ji,
\newblock Phys. Rev. Lett. {\bf 78}, 610 (1997);
%%CITATION = HEP-PH 9603249;%%
A.~Radyushkin,
\newblock Phys. Lett. {\bf B380}, 417 (1996).
%%CITATION = HEP-PH 9604317;%%

\bibitem{Adletal01Chekanov:2003yaAktas:2005ty}
{C. Adloff {\it et al.} (H1 Coll.)},
\newblock Phys. Lett. {\bf B517}, 47 (2001);
%%CITATION = HEP-EX 0107005;%%
ZEUS, S.~Chekanov {\em et~al.},
\newblock Phys. Lett. {\bf B573}, 46 (2003);
%%CITATION = HEP-EX 0305028;%%
H1, A.~Aktas {\em et~al.},
\newblock Eur. Phys. J. {\bf C44}, 1 (2005).
%%CITATION = HEP-EX 0505061;%%

\bibitem{Donnachie:1984xqDonnachie:1985izDonnachie:1987pu}
A.~Donnachie and P.~V.~Landshoff,
\newblock Nucl. Phys. {\bf B244}, 322 (1984); 
%%CITATION = NUPHA,B244,322;%%
\newblock {\em ibid.} {\bf B267}, 690 (1986). 
%%CITATION = NUPHA,B267,690;%%

\bibitem{Donnachie:1998gmDonnachie:2001xx}
A.~Donnachie and P.~V.~Landshoff,
\newblock Phys. Lett. {\bf B437}, 408 (1998); 
%%CITATION = HEP-PH 9806344;%%
\newblock {\em ibid.} {\bf B518}, 63 (2001).
%%CITATION = HEP-PH 0105088;%%

\bibitem{BFKL}
I.~I.~Balitsky and L.~N.~Lipatov,
\newblock Sov.\ J.\ Nucl.\ Phys.\  {\bf 28},  822 (1978); 
%[Yad.\ Fiz.\  {\bf 28} (1978) 1597]
%%CITATION = SJNCA,28,822;%%
E.~A.~Kuraev, L.~N.~Lipatov and V.~S.~Fadin,
\newblock  Sov.\ Phys.\ JETP {\bf 45}  199 (1977).
% [Zh.\ Eksp.\ Teor.\ Fiz.\  {\bf 72} (1977) 377].
%%CITATION = SPHJA,45,199;%%


\bibitem{GluReyVog90}
M.~Gl{\"u}ck, E.~Reya and A.~Vogt,
\newblock Z. Phys. {\bf C48}, 471 (1990);
%%CITATION = ZEPYA,C48,471;%%
%\bibitem{GluReyVog98}
%M.~Gl{\"u}ck, E.~Reya and A.~Vogt,
\newblock Eur. Phys. J. {\bf C5}, 461 (1998).
%%CITATION=Hep-Ph/9806404;%%

\bibitem{Ball:1994kc}
R.~D. Ball and S.~Forte,
\newblock Phys. Lett. {\bf B336}, 77 (1994).
%%CITATION = HEP-PH 9406385;%%

\bibitem{Kotikov:1998qt}
A.~V. Kotikov and G.~Parente,
\newblock Nucl. Phys. {\bf B549}, 242 (1999).
%%CITATION = HEP-PH 9807249;%%

\bibitem{Aid:1995bzChekanov:2001gw}
H1, S.~Aid {\em et~al.},
\newblock Z. Phys. {\bf C69}, 27 (1995);
%%CITATION = HEP-EX 9509001;%%
ZEUS, S.~Chekanov {\em et~al.},
\newblock Nucl. Phys. {\bf B627}, 3 (2002).
%%CITATION = HEP-EX 0202034;%%

\bibitem{BelMueKir01}
A.~V. Belitsky, D.~M{\"u}ller and A.~Kirchner,
\newblock Nucl. Phys. {\bf B629}, 323 (2002).
%%CITATION = HEP-PH 0112108;%%

\bibitem{footnote}
Note that we define $Q^2=-(q_1+q_2)^2/4$, which coincides 
with ${\cal Q}^2=-q_1^2$ only in the forward case.

\bibitem{MueSch05}
D.~M{\"u}ller and A.~Sch{\"a}fer,
\newblock Nucl. Phys. {\bf B739}, 1 (2006).
%%CITATION = HEP-PH 05092004;%%

\bibitem{ManKirSch05}
M.~Kirch, A.~Manashov and A.~Sch{\"a}fer,
\newblock Phys. Rev. Lett. {\bf 95}, 012002 (2005); 
%%CITATION = HEP-PH 0503109;%%
%\bibitem{KirManSch05a}
%M.~Kirch, A.~Manashov and A.~Schafer,
\newblock Phys. Rev. {\bf D72}, 114006 (2005).
%%CITATION = HEP-PH 0509330;%%

\bibitem{IngSch85}
G.~Ingelman and P.~E. Schlein,
\newblock Phys. Lett. {\bf B152}, 256 (1985).
%%CITATION = PHLTA,B152,256;%%

\bibitem{Derrick:1995tw}
ZEUS, M.~Derrick {\em et~al.},
\newblock Phys. Lett. {\bf B356}, 129 (1995).
%%CITATION = HEP-EX 9506009;%%

\bibitem{BelMueNieSch99}
A.~V.~Belitsky, D.~M{\"u}ller, L.~Niedermeier and A.~Sch{\"a}fer,
\newblock Phys. Lett. {\bf B474}, 163 (2000).
%%CITATION = HEP-PH 9908337;%%

\bibitem{FreMcDStr02}
A.~Freund, M.~McDermott and M.~Strikman,
\newblock Phys. Rev. {\bf D67}, 036001 (2003).
%%CITATION = HEP-PH 0208160;%%


\bibitem{Mue05aKumMueKumPasSch06}
D.~M{\"u}ller,
\newblock Phys. Lett. {\bf B634}, 227 (2006);
%%CITATION = HEP-PH 0510109;%%
K.~Kumeri{\v c}ki, D.~M{\"u}ller, K.~Passek-Kumeri{\v c}ki and A.~Sch{\"a}fer,
\newblock in preparation.

\bibitem{Bur0002}
M.~Burkardt,
\newblock Phys. Rev. {\bf D62}, 071503 (2000); 
%%CITATION = HEP-PH 071503%%;
\newblock Int. J. Mod. Phys. {\bf A18}, 173 (2003);
%%CITATION = HEP-PH 0207047;%%
A.~V.~Belitsky, D.~M{\"u}ller, 
\newblock Nucl.\ Phys.\ A {\bf 711},  118 (2002).
%%CITATION = HEP-PH 0206306;%%


\bibitem{Adletal99}
{C. Adloff {\it et. al} (H1 Coll.)},
\newblock Eur. Phys. J. {\bf C13}, 371 (2000).
%%CITATION = HEP-EX 9902019;%%

\end{thebibliography}
%\bibliographystyle{h-physrev4}
%%%%%%%%%%%%%%%%%%%%%%%%%%%%%%%%%%%%%%%%%%%%%%%%%%%%%%%%%%%%%%%%%

%\begin{thebibliography}{99}
%\end{thebibliography}

\end{document}